\documentclass[9pt, a4paper, twocolumn]{article}
\usepackage[left=1.65cm, right=1.65cm, top=1.9cm, bottom=4cm]{geometry}
\usepackage{amsmath}
\DeclareMathOperator*{\argmax}{argmax}
\usepackage{CJKutf8}
\usepackage{authblk}
\usepackage{graphicx}
\usepackage[numbers]{natbib}

\usepackage[utf8]{inputenc}

\makeatletter
\def\thanks#1{\protected@xdef\@thanks{\@thanks
        \protect\footnotetext{#1}}}
\makeatother

\begin{document}

\title{\fontsize{16pt}{1}\selectfont\textbf{Federated Deep Q-Learning and 5G load balancing}\vspace{-0.5em}}
\date{\vspace{-5ex}}

%\begin{CJK*}{UTF8}{bsmi}
\author{\begin{CJK*}{UTF8}{bsmi} 林昕 \end{CJK*}}
\author{\begin{CJK*}{UTF8}{bsmi} 蘇逸康 \end{CJK*}}
\author{\begin{CJK*}{UTF8}{bsmi} 陳竑齊 \end{CJK*}}
\author{\begin{CJK*}{UTF8}{bsmi} 柯拉飛\textsuperscript{*} \end{CJK*}}

%{2}\selectfont\fontsize{11pt}\affil[ ]{
\affil{
\begin{CJK*}{UTF8}{bsmi}
國立臺灣科技大學資訊工程系: 網路協定與資源分配最佳化設計實驗室
\end{CJK*}
\authorcr\{B10732006, B10732026, B10732031,rkaliski\}@mail.ntust.edu.tw%\}
}

\thanks{
\begin{CJK*}{UTF8}{bsmi}
發表於2022年台灣電信年會, 目前版本修改於1月9日2024年
\end{CJK*}
}

%\affil{}
\thanks{
\begin{CJK*}{UTF8}{bsmi}
\textsuperscript{*}通訊作者
\end{CJK*}
}

\thanks{%Rafael Kaliski is thankful for funding support from Taiwan’s Ministry of Science and Technology (MOST), grant:
\begin{CJK*}{UTF8}{bsmi}
本研究由科技部贊助，計畫編號MOST 108-2218-E-011-036-MY3
\end{CJK*}
}

%===============================================================
%\end{CJK*}
%===============================================================

\maketitle 
\pagestyle{empty}
\thispagestyle{empty}
\setlength{\parindent}{2em}
\setlength{\affilsep}{1mm}
%%%%%%%%%%%%%%%%%%%%%%%%%%%%%%%%%%%%%%%%%%%%%%%%%%%%%%%%%%%%%%%%%%%%%
%%%%%%%%%%%%%%%%%%%%%%%%%%%%%%%%%%%%%%%%%%%%%%%%%%%%%%%%%%%%%%%%%%%%%
\begin{CJK*}{UTF8}{bsmi}
\begin{abstract} 
儘管蜂窩網絡技術取得了進步，但基站的負載平衡仍然是一個長期存在的問題。雖然集中式資源分配方法可以解決負載平衡問題，但它們仍然是個 NP-hard 的問題且對於實務上的網絡部署並不實際。在這項實驗中，我們研究如何使用 federated deep Q learning來通知每個用戶設備 (UE) 每個基站(BS)的負載情況。 Federated deep Q learning 的負載均衡可以使智能 UE 能夠獨立選擇最佳基站，同時也限制暴露給網絡的私人信息量。

在這項研究中，我們提出並分析了一個federated deep Q learning 的負載平衡系統，此系統利用 Open-RAN xAPP 框架和 near-Real Time Radio Interface Controller（near-RT RIC）來實作。我們的模擬結果指出，相較於 UE 當前所採用的最大化傳輸量的maximum Signal-To-Noise-Ratio (MAX-SINR) 方法相比，我們所提出的 deep Q learning 模型可以始終如一的提供更高的平均 UE 服務品質 (QoS)。
\newline
\textbf{關鍵字：Deep Q-learning，5G，負載平衡，Open RAN}
\end{abstract}

\section{研究背景}
負載平衡在5G以及往後世代的網路扮演的角色越來越重要，隨著更高傳輸，低延遲的應用，如高品質影音、虛擬實境、無人車網路的出現，UE對於位元速率的要求越來越嚴苛，即便使用了高頻段的毫米波來提升傳輸效率，也會因為大量的設備連接導致無線網路資源匱乏，設備所能得到的資源塊數量過少。因此有很多論文都在探討如何能分散各個基地台的流量，避免太多設備同時競爭，達到使用效率的最大化。另外由ORAN Alliance所提出的Open RAN\cite{O_RAN}架構中，不僅是制定了多個可交換訊息的介面，方便資料的採集及分析，也在機器學習模型的部屬上提供了高度的支援，相較於傳統上用於負載平衡的方法，如MAX-SINR，機器學習能夠更好的分析各個基地台的負載狀況，為各個UE選擇最適合連接的基地台。

然而現今提出的相關論文很多都是以集中式學習為模型架構。雖然集中式學習模型可以很有效的做到負載分析，但其考慮整體環境狀況的模式在實際部屬上的可行性還有待確認，尤其模型的複雜度會相當高，增加了伺服器的運算壓力，決策效率也會因此降低而無法及時反映網路的變化。此外，UE必須將其本身的網路狀態、位置、資料要求量等等傳輸到伺服器才能做決策，這可能會增加有心人士盜取設備訊息的機會。為了避免上述情況發生，在這篇論文中我們提出了我們的方法，每個在網路環境內的UE接收由伺服器送出的基地台負載資訊，結合自身的狀況，透過部屬於自身內部的小型模型做分析，以最大化自身傳輸速率為目標選出最適合自己的基地台並將分析結果送到伺服器。伺服器除了每隔一段時間要送出各個基地台的負載資訊給各個UE之外，接收到各個UE的分析結果後，伺服器也要負責統整確認各個UE的決策都可行，並指揮基地台執行各個UE的換手需求。相較於集中式學習，我們的分析模型因為只有考慮了UE自身的狀態和基地台負載資訊而非對整個網路做管理，複雜度可以大幅度的降低，且因為各個UE是獨立做決策，可以有效利用各個UE的運算能力而不需完全仰賴伺服器的運算資源，綜合起來不僅能提升整體的決策效率，也能夠釋放伺服器的運算壓力。在UE和伺服器的資料交換上僅有各個基地台的負載以及UE的決策，對於UE的私人資訊也能提供有效的保護。

這篇論文後續由以下章節組成，第二章列出了我們所參考的相關研究，簡單描述了其內容以及對於這篇論文的影響，第三章描述了Open ran的大致架構，包含各個元件的介紹以及對於機器學習模型部屬上的幫助，第四章為我們的實作方法，對我們使用的模型架構以及資料傳遞的流程做詳細說明，最後第五章為我們使用的模擬環境設置以及實驗結果分析。

\section{相關研究}
基地台之間的連接性和負載平衡仍然是一個難以解決的問題。如今有許多研究嘗試以各種不同的模型架構如集中式學習、分散式學習、multi agent等等來嘗試解決。\cite{9354748},\cite{8057766}使用了集中式的Deep Q-learning模型做負載平衡，相較於Q-learning，Deep Q-learning較不受限制，能對具連續性的資料進行分析。在實驗結果上比起Q-learning有更快的收斂速度以及更好的表現。這兩篇論文提供我們模型架構的參考。\cite{9303466}在Q learning的基礎上使用了Transfer learning的概念，在Resource task和Target task同時訓練不同的model，並使用對照函式將次要任務轉換至主要任務，藉此Target task可以得到更多用於訓練的資料。實驗結果顯示模型使用了Transfer learning在收斂效率和reward上表現得更好。\cite{mollel2020multi}提出了將CNN和Multi-agent Deep Q learning結合的基地台選擇策略，這篇論文將策略分為兩個階段，首先將網路分布狀況利用CNN分析找尋其中的特徵如基地台位置、UE位置，接著將分析出的特徵輸入進各個agent做決策，以最大化傳輸量為目標的同時進行負載平衡。實驗結果顯示換手次數相較於現今常用的方法有顯著的下降，但這篇論文並沒有提供傳輸量相關的比較。

\section{Open RAN介紹}
\begin{figure}[t]
    \centering
    \includegraphics[scale=0.5]{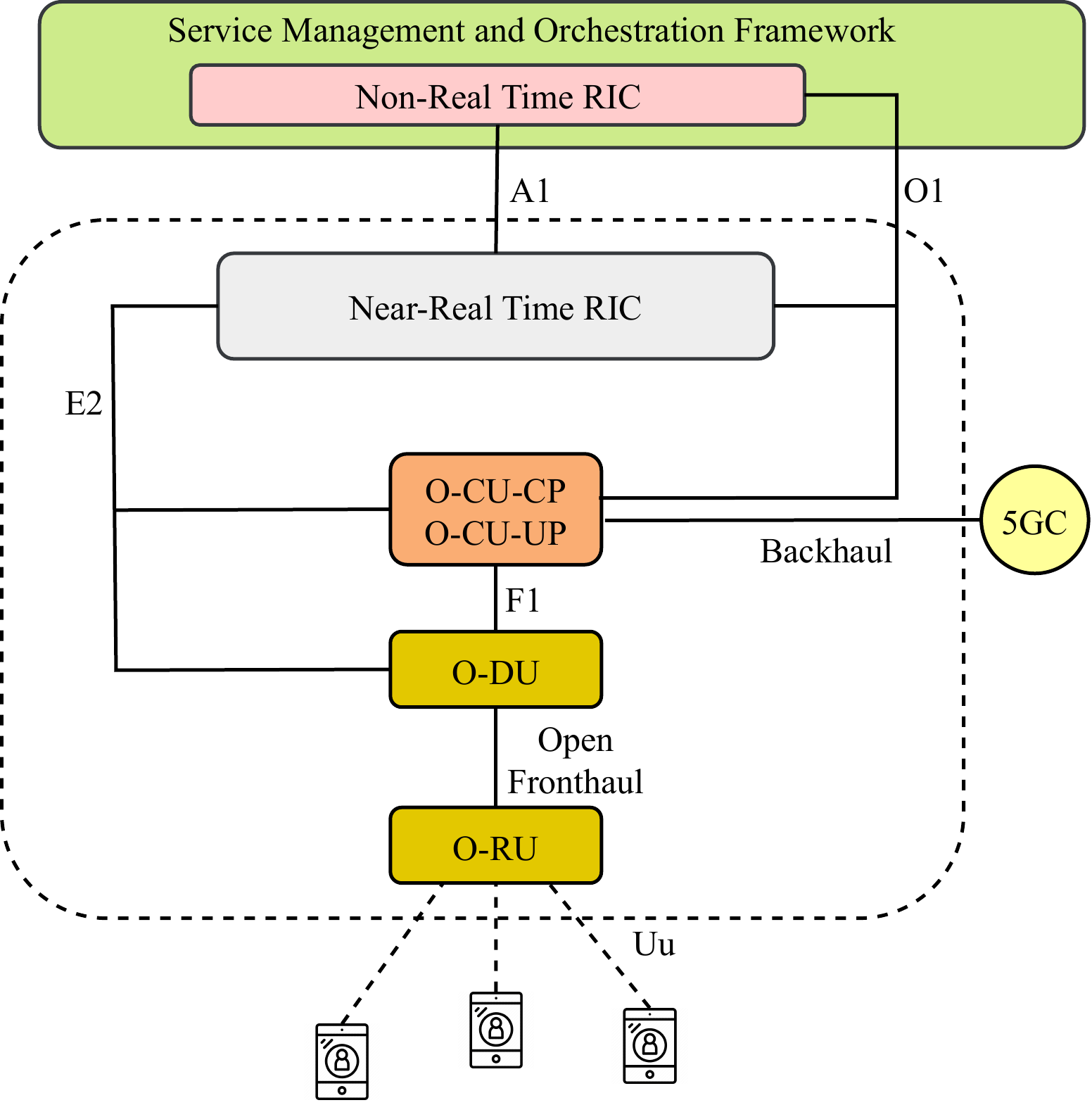}
    \caption{O-RAN架構}
    \label{fig:1}
\end{figure}
傳統的Radio Access Network(RAN)通常是各家企業獨立開發，許多元件都以達成特定目的做設計，無法為使用者提供足夠的彈性，並且各家企業的開發內容並沒有共通的標準介面互相傳遞訊息，也提升了機器學習部屬的難度。為此O-RAN Alliance提出了Open RAN構想，透過軟體定義網路（SDN, Software Defined Network）來制定各個元件及交換訊息用的標準化介面，並提供了相當大的部屬彈性，讓使用者可以根據自己的需求部屬需要的功能。圖\ref{fig:1}簡單描述了Open RAN\cite{Parallel_wireless, O_RAN}的大致架構和各個元件互相溝通所使用的介面，以下簡單介紹其中較為重要的元件及溝通介面：
\begin{enumerate}
    \item SMO(Service Management and Orchestration)：
    
    藉由O1介面接收整個網路的運作資訊，並提供網路設施的管理服務，監控與其連接之元件的資源消耗以及效能。
    
    \item Non RT RIC：
    
    部屬於Non RT RIC的rapp會透過O1所提供的資訊，進行資料的分析以及模型訓練，並經由A1介面將訓練好的模型或策略傳到Near RT RIC平台，將其投入應用。也可以透過A1介面監控模型及策略的運行狀況，以及管理部屬於Near RT RIC平台的xapp。通常而言運作週期在1秒以上。
    
    \item Near RT RIC：
    
    透過E2並接收各個元件的運作情況，進行分析預測後，同樣透過E2介面下達決策來調整個網路的行為。另外各個元件的狀態會儲存在Near RT RIC內部的SDL(Shared Data Layer)資料庫中，供各個xApp讀取。運作週期在10毫秒到1秒之間。
    
    \item xApp：
    
    部屬在Near RT RIC內，從E2介面接收網路狀態後進行分析，並下達決策來調整網路行為。ORAN Alliance提供了相當大的彈性，使用者可以自己設計xApp並部屬於Near RT RIC，來最佳化自己的網路設施。
    
    \item O-CU(Centralized unit)：
    
    處理上層的無線電操作，例如Radio Resource Control (RRC)、Packet Data Convergence Protocol (PDCP)和Service Data Application Protocol (SDAP)。

    \item O-DU(Distributed unit)：
    
    主要運行如Radio Link Control (RLC)、Medium Access Control(MAC)以及部分PHY層的操作。

    \item O-RU(Radio unit)：
    
    處理fronthaul的操作，如各項PHY層動作、數位波束成形以及Fast Fourier Transform (FFT)。

\end{enumerate}

\section{提出方法}
\begin{figure}[t]
    \centering
    \includegraphics[scale=0.4]{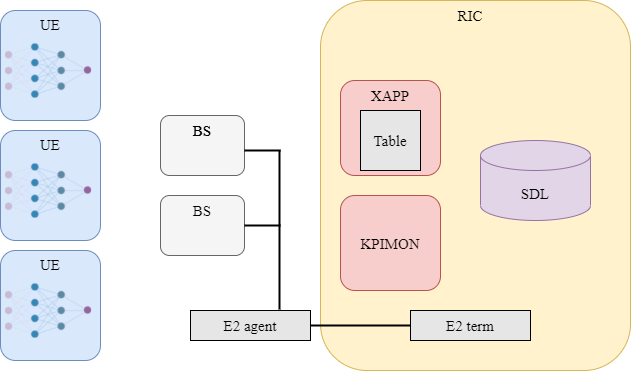}
    \caption{提出之架構}
    \label{fig:2}
\end{figure}
在這篇論文中，如圖\ref{fig:2}所示，我們提出每個本地端運行各自的Deep Q Network模型的方法，讓各個UE利用自身的模型決策出對自己最有利的基地台換手策略，並透過O-RAN的Near Real-Time RIC平台中的xApp進行換手策略整理，再將使用者裝置進行換手，資料傳輸一段時間後，模型便會再次評估目前的狀態，並進行下一次決策。以下詳細說明我們的實作方法:
\begin{enumerate}
    \item 本地端的模型
    
\begin{figure}[t]
    \includegraphics[trim=0mm 0mm 20mm 55mm,clip,width=\columnwidth]{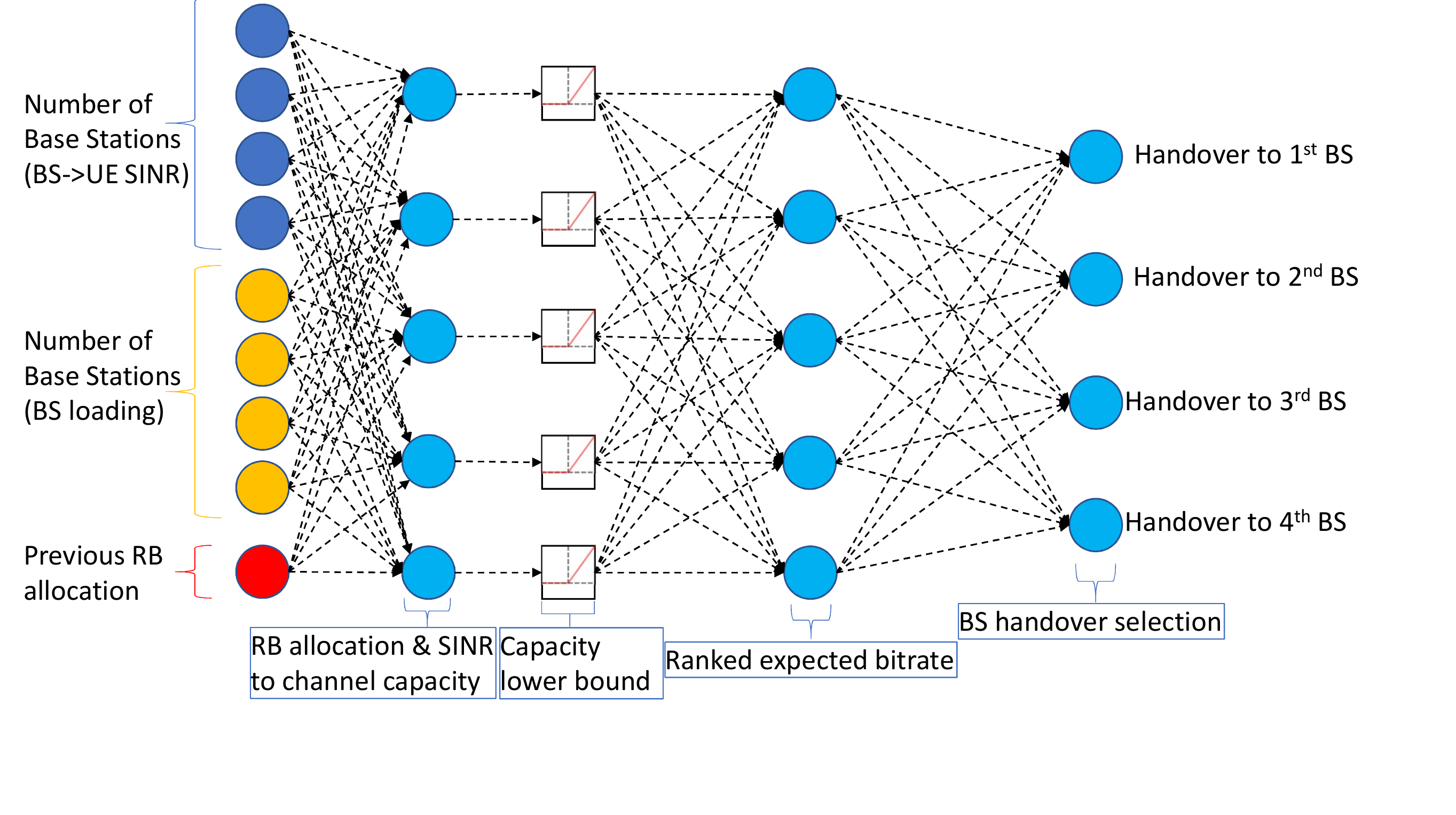}
    \caption{神經網路架構} %，因為各基地台的資源塊數量都一直，不用另外放進去
    \label{fig:4}
\end{figure}
    我們定義在一個網路環境中，基地台集合為$\{{BS}_1, {BS}_2, ... ,{BS}_m\}$其中m為此網路環境中的基地台總數；本地端設備集合為 $\{{UE}_1, {UE}_2, ... ,{UE}_n\}$其中n為本地端設備總數。在每一個本地設備中，我們部屬了一個五層的神經網路，如圖\ref{fig:4}所示，其中包含了輸入及輸出層，以及三個隱藏層。每個UE的神經網路參數定義為$\theta_i$。在state，也就是神經網路的輸入方面，我們定義如(1)所示：
\begin{equation}
    S_i = \{\rho^i_1, \rho^i_2, ..., \rho^i_m, l\rho_i, l_1, l_2,...,l_m\}, 1\le i \le n
\end{equation}
    
    其中$S_i$表示${UE}_i$的state；$\rho^i_j$表示${UE}_i$和${BS}_j$之間的SINR；${lp}_i$表示${UE}_i$得到的資源塊數量， $l_j$表示${BS}_j$的負載量。而SINR的定義如(2)所示：

\begin{equation}
    \rho^i_j = \frac{rxp(i,j)}{\sum\limits^m_{x=1,x\ne j} rxp(i, x) + \sigma}, 1\le i\le n, 1\le j \le m
\end{equation}
    
    $rxp(i, j)$表示${UE}_i$從${BS}_j$收到的訊號強度，表示環境的路徑傳輸損失以及雜訊。在action $A_i$，也就是神經網路的輸出，同時也是設備要採取的行為方面，我們定義如(3)所示：
    
\begin{equation}
    A_i = \{a^i_1, a^i_2,... ,a^i_m\}, 1\le i \le n
\end{equation}
    
    其中$a^i_j$表示將${UE}_i$連接至${BS}_j$。而在reward $r^i$方面則是定義為設備所得到的位元速率。我們採用了$\epsilon - greedy$作為選擇行為的策略，各個設備中的模型將這個設備當前的state輸入神經網路後，會有$\epsilon$的機率以下列方式選擇action $a^{*i}$：

\begin{equation}
    a^{*i} = \argmax_{a^i_j \in A_i} Q( S_i, a^i_j | \theta_i)
\end{equation}

    其中$Q( S_i, a^i_j | \theta_i)$表示在$\theta_i$這組參數下以$S_i$作為輸入，$a^i_j$所對應的Q value。另外有$1-\epsilon$的機率模型會隨機選擇一個action，這樣的方式可以提升模型在各個情境下的探索能力，避免收斂在比較差的決策上。而在模型得到來自環境的reward之後，便以下列方式做模型的更新：

\begin{equation}
    Q(S_i, a^{*i} | \theta^{'}_{i}) = r^i + \gamma \max \left( Q(S^{'}_{i}, a^i_j | \theta^{*}_{i}) \right)
\end{equation}

    $\gamma$為一個折扣係數， $S^{'}_{i}$表示在執行action之後進入的state， $\theta^{*}_{i}$表示目標模型參數。更新的原理是基於各action的目標Q value及當前Q value存在的誤差，並藉由後傳來調整神經網路的參數。
    
    \item Near RT RIC xapp：
    
    在Near RT RIC平台上我們部屬了一個xapp，負責管理這個網路環境中所有基地台的負載情況。xapp會不斷地向E2 agent要求最新的環境資訊，而各個UE決定要採取的action後，也會將決策經由E2 agent傳遞至Near RT RIC，當xapp收到UE的決策，會先比對自身紀錄的基地台負載資訊，確認無誤之後向指定的基地台發出換手指令。每經過一個時間段，xapp會以廣播的方式告知所有UE最新的基地台負載情況，各個UE藉此更新自身的state，來進行下一次的決策。
\end{enumerate}

\section{實驗結果}

在這篇論文中，我們模擬了一個簡單的UE、基地台互動環境，採用urban和Non-Line Of Sight（NLOS）的模式，而基地台的種類採用了macro cell。%為了降低運算難度，我們選擇不考慮shadow、fading和interference，
為了降低運算難度以及評估長時間的訊號品質，我們選擇不考慮shadowing、fading和interference，而各基地台的訊號不會干擾到其他基地台，因此使用者裝置與基地台的連線品質完全取決於由雙方的距離以及基地台的負載情況。在表格\ref{tbl:Sim_Params}中的參數是我們使用的模擬環境參數，其中的round robin為一個傳輸時間間隔（TTI）中（1ms），一個基地台一次服務10個UE，服務後的UE再排到服務序列尾端，等待下一次的服務。

{\fontsize{8pt}{3}\selectfont\begin{table}[ht]
	\centering
	\caption{模擬環境參數}
	\label{tbl:Sim_Params}
	\begin{tabular}{|l|l|l|} 
		%\backslashbox{Type}{Target Audience} & Low & High \\
		\hline \textbf{參數名稱} & \textbf{數值} \\
		\hline 區域 & 2公里 (長) x 1.5公里 (寬) \\
		\hline 基地台數量 & 4 \\
		\hline Path loss 模式 & UMA NLOS\cite{TR_38901} \\ %TR37.876?
		\hline 雜訊 & -95dBm \\
		\hline 基地台頻寬 & \{50, 100\} \\
		\hline 使用者裝置數量 & \{20, 40, 50, 60, 80, 100\} \\
		\hline 資源分配方式 & Round Robin (Max Throughput)\\
		\hline 載波頻率 & 3.5 GHz \\
		\hline 基地台天線高度 & 25m \\
		\hline 使用者天線高度 & 1.5m \\
		\hline 模擬時長 & 模擬24小時, 數據以每分鐘為單位 \\
		\hline
	\end{tabular}
\end{table}}
%Note:  We may need to add SINR-RB capacity mapping

\begin{figure}[t]
    \centering
    \includegraphics[scale=0.7]{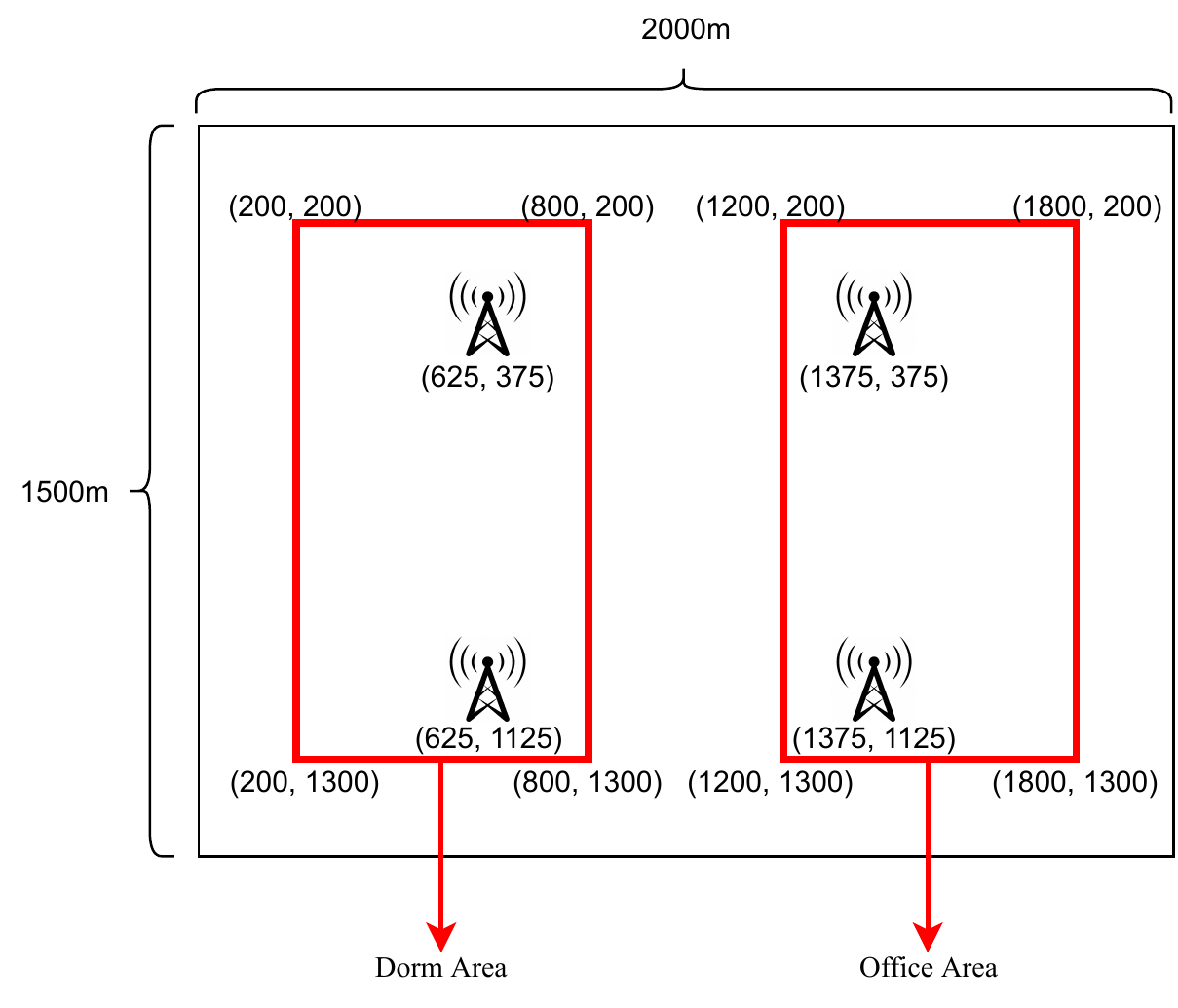}
    \caption{環境示意圖}
    \label{fig:3}
\end{figure}
整體環境位置的分布如圖\ref{fig:3}所示，四個macro cell的位置平均分布在地圖上。地圖主要劃分成兩個區域，分別為居住區以及辦公區，而使用者裝置則是均勻地分散在這兩個區域。隨著時間的推進，兩邊的使用者會以固定速度朝另一區域前進，模擬出上班以及下班的狀況。我們模擬了各個裝置在一天內的行為，為了加速模擬過程，模擬環境將現實的1秒鐘視為環境中的1分鐘，因此一次模擬需花費實際時間24分鐘，每個使用者裝置總共有1440筆的數據，其中每筆數據包含當前位置以及其要求的資料量。

然而礙於設備以及軟體限制，我們無法一次同時模擬太大量具有運算能力的使用者裝置，因此我們將所有的模型都放到我們自行編寫的xApp上，來觀察模型在不同環境設置下的表現。如圖\ref{fig:2}所示，模擬器運行出來的數據，透過e2agent傳遞給另一個名為kpimon的xApp，kpimon將其寫入SDL，我們自行編寫的xApp便可讀取SDL來獲取資料和對其進行分析及運算，並讓同使用者裝置數量的模型進行預測，最終得出其判定的最佳基地台換手策略。我們所使用的模型參數如表\ref{tbl:model_para}所示。

\begin{figure}[t]
    \centering
    \includegraphics[scale=0.4]{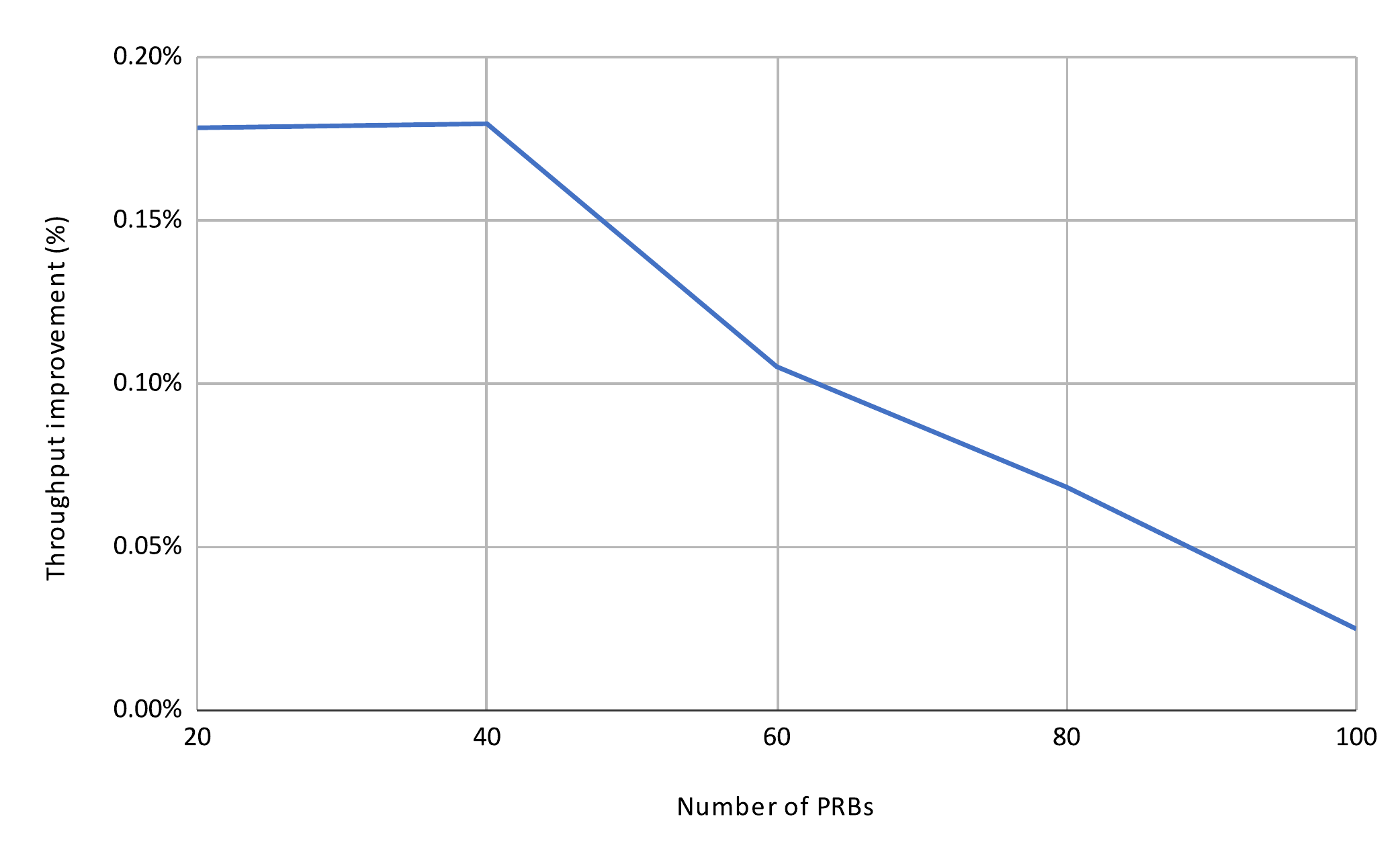}
    \caption{固定UE數量為100，不同頻寬之流量變化}
    \label{fig:fixUE_TH}
\end{figure}

\begin{figure}[t]
    \centering
    \includegraphics[scale=0.4]{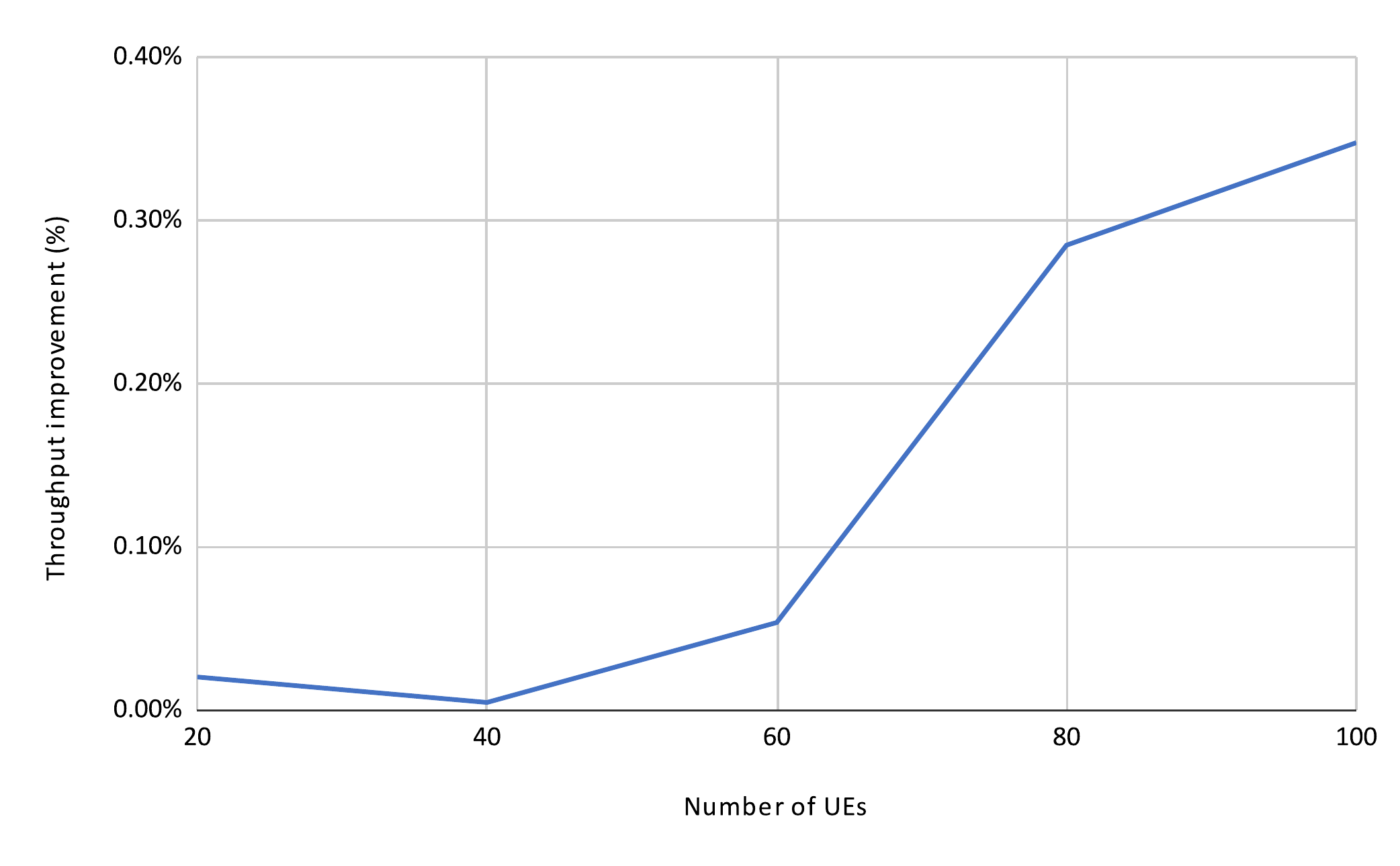}
    \caption{固定頻寬為50RB，不同人數之流量變化}
    \label{fig:fixPRB_TH}
\end{figure}

\begin{figure}[t]
    \centering
    \includegraphics[scale=0.4]{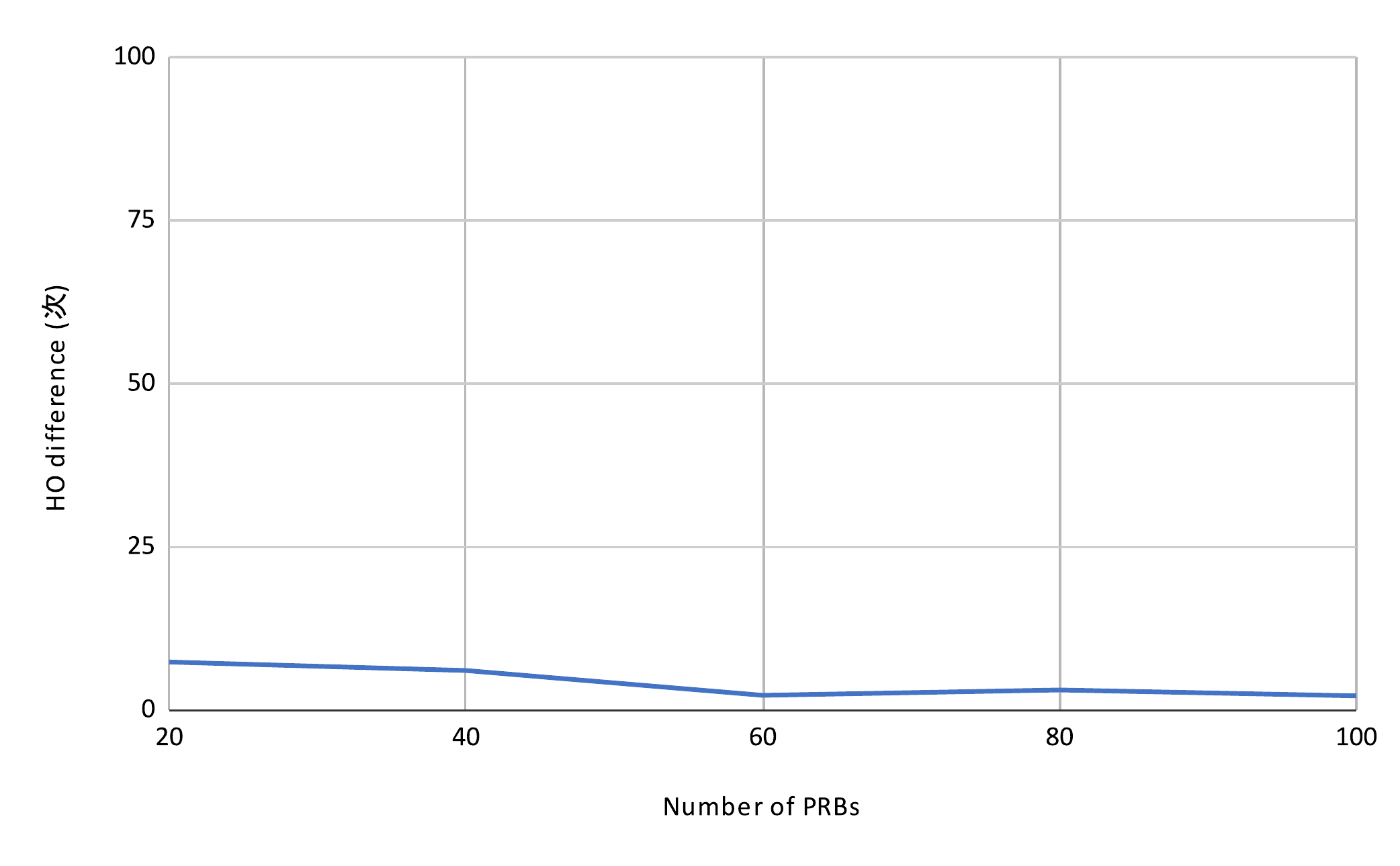}
    \caption{固定UE數量100，不同頻寬之換手次數變化}
    \label{fig:fixUE_HO}
\end{figure}

\begin{figure}[t]
    \centering
    \includegraphics[scale=0.4]{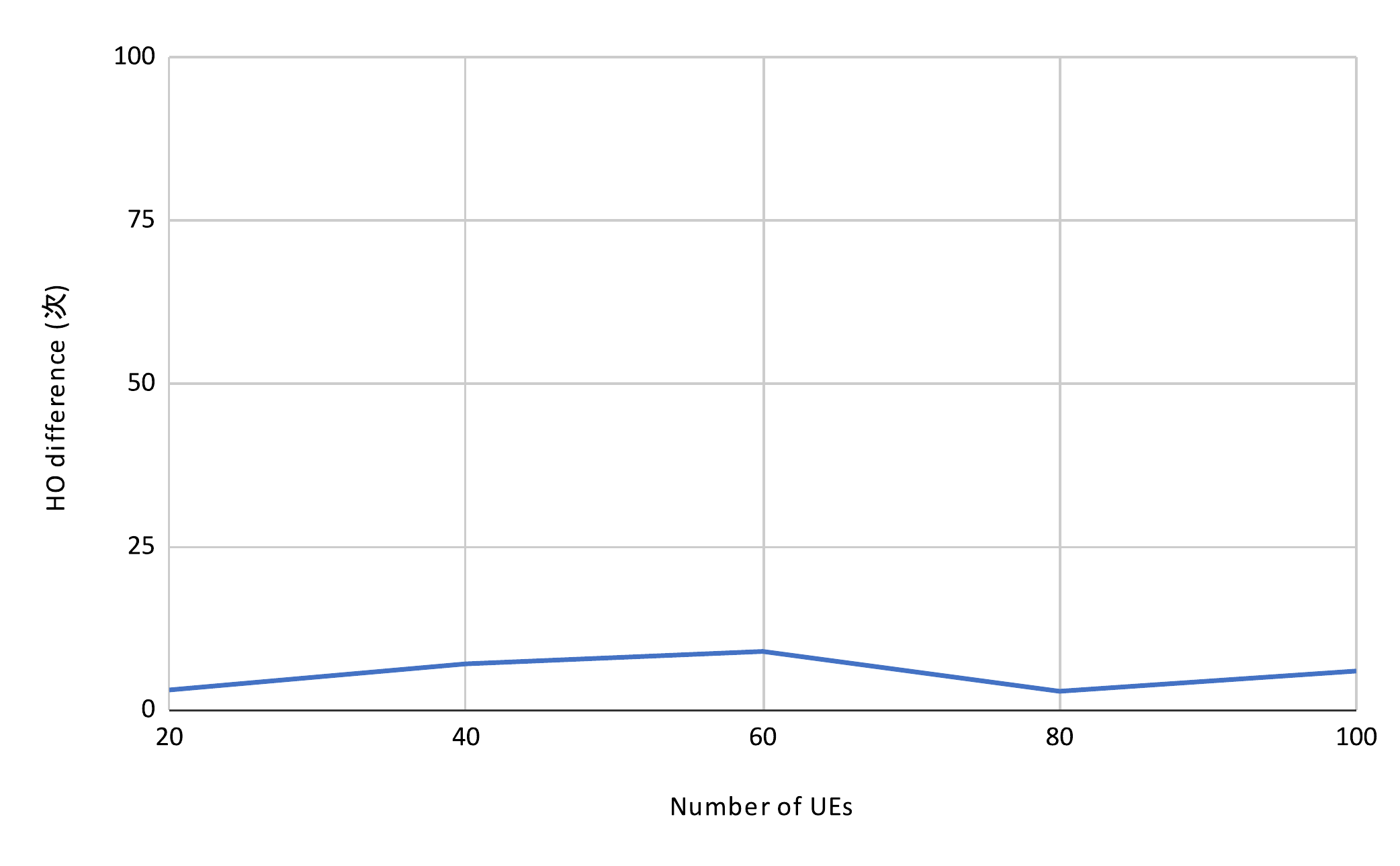}
    \caption{固定頻寬為50RB，不同人數之換手次數變化}
    \label{fig:fixPRB_HO}
\end{figure}

{\fontsize{8pt}{3}\selectfont\begin{table}[ht]
	\centering
	\caption{機器學習模型參數}
	\begin{tabular}{|l|l|l|} 
		%\backslashbox{Type}{Target Audience} & Low & High \\
		\hline \textbf{參數} & \textbf{數值} \\
		\hline 學習率 & 0.01 \\
		\hline Discount factor ($\gamma$) & 0.9 \\
		\hline Memory capacity & 480 \\ %TR37.876?
		\hline Batch 大小 & 150 \\
		\hline Action epsilon ($\epsilon$) & 0.8 \\
		\hline Episodes次數 & 100 \\
		\hline
	\end{tabular}
	\label{tbl:model_para}
\end{table}
}
我們將我們的換手演算法與MAX-SINR進行比較。MAX-SINR的方式為UE單純挑選周遭基地台中SINR值最高的基地台進行連線，而沒有考慮基地台的負載情況。我們比較的依據為換手次數與總體流量，而以下有不同的測試變數：
    \begin{enumerate}
        \item 固定UE數量，觀察改變基地台頻寬後的變化, 圖\ref{fig:fixUE_TH},圖\ref{fig:fixUE_HO}
        \item 固定基地台頻寬，觀察改變UE數量後的變化, 圖\ref{fig:fixPRB_TH},圖\ref{fig:fixPRB_HO}
    \end{enumerate}
	
首先是流量的變化，在第一種測試變數下，如圖\ref{fig:fixUE_TH}所示，X軸為基地台提供不同頻寬的情況，可以看出隨著PRB數量增加，可用資源變多導致執行負載平衡的效益愈發不明顯；而在第二種測試變數下，如圖\ref{fig:fixPRB_TH}所示，X軸為不同的UE數量情況，雖然整體流量提升的幅度不大，但平均的QoS有所提升，且隨著UE數量增加，資源的競爭更加激烈，負載平衡的效益便凸顯出來。由此可以發現，我們考慮基地台負載情況後，的確是可以提升整體的連線流量。

再來是換手次數的差距，在第一種測試變數下，如圖\ref{fig:fixUE_HO}所示，X軸為基地台提供不同頻寬的情況，而在第二種測試變數下，如圖\ref{fig:fixPRB_HO}所示，X軸為不同的UE數量情況，另外兩圖的Y軸則是將使用模型後的總換手次數減去使用MAX-SINR的總換手次數所統計出的次數差距。由結果可以看出，在大部分情況下使用機器學習的方法會造成輕微的換手次數提升，這是因為模型在換手之後可能會對其他UE產生連鎖效應，讓其他UE出現更好的連線選擇。

\section{總結}
對比其他論文的集中式訓練，我們提出在5G環境下採用聯盟式訓練的方法，在每個使用者設備部屬自己本地端的模型，並且透過本地產生出來的數據直接套進模型裡預測，來進行整體環境的換手策略。如此除了可以縮小模型的複雜度以及部屬難度，也可以有效地減輕伺服器運算壓力，和降低資訊安全的疑慮。另一特色是我們也與O-RAN進行整合，透過O-RAN高度支援SDN的特性，將機器學習模型以及一些資料前處理之方程式，一同編寫至我們自己定義的xApp，最後再透過O-RAN的Near Real-Time RIC平台以及5G環境模擬器，進行換手策略的實驗。模型透過加入負載變化來進行預測，相比單純只看訊號強度的策略，在整體流量有些微的提升。往後，我們會持續找出能夠大幅提升流量的演算法，並且與現今機器學習領域較為全新的訓練方式—聯盟式學習（Federated Learning）進行整合。
\end{CJK*}

\bibliographystyle{unsrt}
\bibliography{IEEEabrv,reference}

\begin{CJK*}{UTF8}{bsmi}
\section{附件}
    \begin{enumerate}
        \item 程式碼: https://reurl.cc/akk7XY
        \item 訓練集: https://reurl.cc/veeRvN
    \end{enumerate}
\end{CJK*}
\end{document}